\documentclass{article}
\usepackage{natbib}
\usepackage{amssymb}
\usepackage{amsmath}
\usepackage{graphicx}
\begin{document}

\title{On Stackelberg Mixed Strategies\thanks{This paper appears in {\em
      Synthese}, Volume 193, Issue 3, pp.~689-703, March 2016.  The final
    publication is available at Springer via
    http://dx.doi.org/10.1007/s11229-015-0927-6}}

\author{Vincent Conitzer\\Duke University}
\date{}
\maketitle

\begin{abstract}
  It is sometimes the case that one solution concept in game theory is
  equivalent to applying another solution concept to a modified version of
  the game.  In such cases, does it make sense to study the former
  separately (as it applies to the original representation of the game), or
  should we entirely subordinate it to the latter?  The answer probably
  depends on the particular circumstances, and indeed the literature takes
  different approaches in different cases.  In this article, I consider the
  specific example of Stackelberg mixed strategies.  I argue that, even
  though a Stackelberg mixed strategy can also be seen as a subgame perfect
  Nash equilibrium of a corresponding extensive-form game, there remains
  significant value in studying it separately.  The analysis of this
  special case may have implications for other solution concepts.
\end{abstract}

\newpage

\section{Introduction}

Game theory provides ways of representing strategic situations, as well as
{\em solution concepts} indicating what it means to ``solve'' the resulting
games. These are intertwined: a solution concept may be meaningfully
defined only for some ways of representing games.  Moreover, sometimes, a
solution concept is equivalent to the application of another solution
concept to a transformation of the original game.  In this case, one may
wonder whether it is sensible to study the former concept separately.  One
might well argue that we should only define the latter concept, and see the
former as just an application of it, for the sake of parsimony.  Entities
should not be multiplied unnecessarily!

In this article, I consider the case of {\em Stackelberg mixed strategies},
which are optimal mixed strategies to commit to.  It will be helpful to
first review Stackelberg models in general.  A (two-player) Stackelberg
model involves one player being able to act (or commit to a course of
action) before the other player moves.  The standard example is that of two
firms competing on quantity. If one firm is able to commit to a quantity
before the other moves (Stackelberg competition), the committing firm can
benefit significantly from this in comparison to the model where both firms
move simultaneously (Cournot competition). (For more detail, see,
e.g.,~\cite{Fudenberg91:Game_theory}.) A Stackelberg model requires that
the commitment is absolute: the Stackelberg leader cannot backtrack on her
commitment.  It also requires that the other player, the Stackelberg
follower, sees what the leader committed to before he himself moves.

Of course, we can consider what happens if one player obtains a commitment
advantage in other games as well.\footnote{One line of work concerns
  settings where there are many selfish followers and a single benevolent
  leader, for example a party that ``owns'' the system and controls part of
  the activity in it, who acts to optimize some system-wide objective.
  See, e.g.,~\cite{Roughgarden04:Stackelberg}. In this article I will not
  assume that the leader is benevolent.}  We can take any two-player game
represented in {\em normal form} (i.e., a bimatrix game), and give one
player a commitment advantage.  The game in Figure~\ref{fi:standard_game}
is often used as an example.
\begin{figure}
\begin{center}
{\large
\begin{tabular}{r|c|c|}
 & $L$ & $R$  \\ \hline
$U$ & 1,1  & 3,0  \\ \hline
$D$ & 0,0  & 2,1  \\ \hline
\end{tabular}}
\caption{A game that illustrates the advantage of commitment.}
\end{center}
\label{fi:standard_game}
\end{figure}
In this game, if neither player has a commitment advantage (and so they
make their choices simultaneously), then player $1$ (the row player) has a
{\em strictly dominant strategy}: regardless of player $2$'s choice, $U$
gives player $1$ higher utility than $D$.  Realizing that player $1$ is
best off playing $U$, player $2$ is better off playing $L$ and getting $1$,
rather than playing $R$ and getting $0$.  Hence, $(U,L)$ is the solution of
the game by {\em iterated strict dominance} (also implying that it is the
only equilibrium of the game), resulting in utilities $(1,1)$ for the
players.

Now suppose that player $1$ can {\em commit} to an action (and credibly
communicate the action to which she has committed to player $2$) before
player $2$ moves.  If she commits to playing $U$, player $2$ will again
play $L$.  On the other hand, if she commits to playing $D$, player $2$
will realize he is better off playing $R$.  This would result in utilities
$(2,1)$ for the players.  Hence, player $1$ is now better off than in the
version of the game without commitment.\footnote{Note that player $1$
  merely {\em stating} that she will play $D$, without any commitment, will
  not work: she would always have an incentive to back out and play $U$
  after all, to collect an additional $1$ unit of utility, regardless of
  what player $2$ plays.  Player $2$ will anticipate this and play $L$
  anyway.}

\begin{figure}
\begin{center}
\includegraphics[width=\linewidth, trim=55 200 55 150]{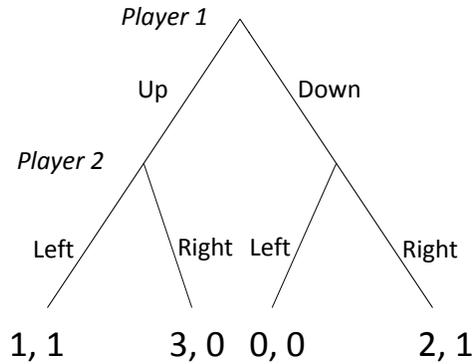}
\end{center}
\vspace{.1in}
\caption{The extensive-form representation of the pure Stackelberg version
  of the game in Figure~\ref{fi:standard_game}.}
\label{fi:standard_extensive}
\end{figure}

While in this example, the Stackelberg outcome of the game is different
from the simultaneous-move outcome, my impression is that most game
theorists would not consider the Stackelberg outcome to correspond to a
different {\em solution concept}.  Rather, they would see it simply as a
different {\em game}.  Specifically, the time and information structure of
the game---who moves when knowing what---is different.  The {\em extensive
  form} provides a natural representation scheme to model the time and
information structure of games.  For example, the Stackelberg version of
the game in Figure~\ref{fi:standard_game} can be represented as the
extensive-form game in Figure~\ref{fi:standard_extensive}.
This game is easily solved by backward induction: if player 1 has committed
to Up, then it is better to move Left for player 2, resulting in utilities
$(1,1)$; on the other hand, if player 1 has committed to Down, then it is
better to move Right for player 2, resulting in utilities $(2,1)$.  Hence,
player 1 is best off moving Down.  Thus, solving the extensive-form game by
backward induction gives us the Stackelberg solution.

So far, so reasonable.  Now, let us turn to Stackelberg {\em mixed}
strategies.  Here, one of the players has an even stronger commitment
advantage: not only is she able to commit to a course of action, she is
able to commit to a mixed strategy, that is, a {\em distribution} over the
actions that she can take.  Consider again the game from
Figure~\ref{fi:standard_game}, and now suppose that player $1$ can commit
to a mixed strategy.  She could commit to the distribution $(0,1)$, i.e.,
putting all the probability on $D$, and again obtain $2$.  However, she can
do even better: if she commits to $(0.49,0.51)$, i.e., putting slightly
more than half of the probability mass on $D$, player $2$ will still be
better off playing $R$ (which would give him $1$ slightly more than half
the time) than playing $L$ (which would give him $1$ slightly less than
half the time).  This results in an expected utility of
$0.49 \cdot 3 + 0.51 \cdot 2 = 2.49 > 2$ for player $1$.  Of course, player
$1$ can also commit to $(0.499,0.501)$, and so on; in the limit, player $1$
can obtain $2.5$.  Stackelberg mixed strategies have recently received
significant attention due to their direct application in a number of real
security
domains~\citep{Pita08:Using,Tsai09:IRIS,An12:PROTECT,Yin12:TRUSTSAIMAG}.

\begin{figure}
\begin{center}
\includegraphics[width=\linewidth, trim=55 160 55 190]{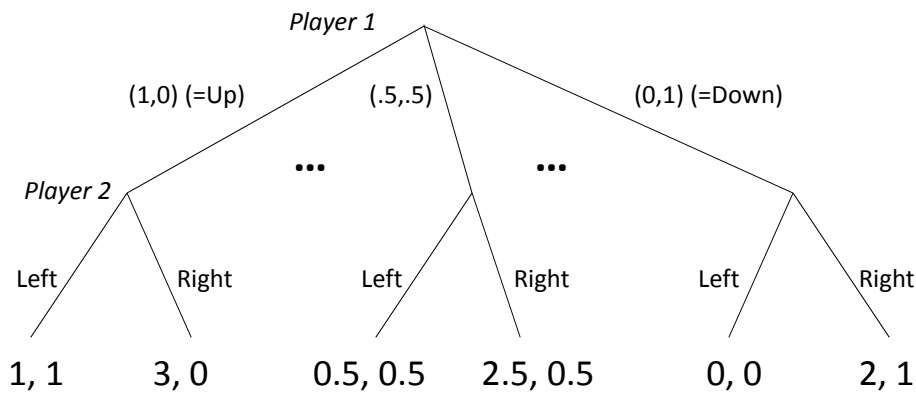}
\end{center}
\vspace{.1in}
\caption{The extensive-form representation of the mixed Stackelberg version
  of the game in Figure~\ref{fi:standard_game}.}
\label{fi:mixed_extensive}
\end{figure}

Again, it is possible to capture the commitment advantage that player $1$
has using the extensive form, as illustrated in
Figure~\ref{fi:mixed_extensive}.
Note that player $1$ has a continuum of moves in the first round, as
indicated by the ellipses.  Each of the (infinitely many) subgames has a
straightforward solution, with the exception of the one where player $1$
has committed to $(0.5,0.5)$, in which player $2$ is indifferent between
his choices.  If player $2$ responds by playing Right in this case, then it
is optimal for player $1$ to in fact commit to $(0.5,0.5)$; and this
constitutes the unique subgame perfect Nash equilibrium of the game.

Again, this way of representing the game in extensive form and solving it
is entirely correct.  However, it appears more awkward than it did in the
case of committing to a pure strategy.  For one, the first node in the game
tree now has infinitely many children.  This is due to the fact that
committing to a mixed strategy is {\em not} the same as randomizing over
which pure strategy to commit to.  The reason that they are not the same is
that if player $1$ randomizes over which pure strategy to commit to, then
player $2$ sees the {\em realization} of that random process, i.e., the
realized pure strategy, before acting.  Because of this, there is indeed no
reason to randomize over which pure strategy to commit to, as this could
not lead to a higher utility than simply committing (deterministically) to
whichever pure strategy maximizes player $1$'s utility. Consequently,
randomizing over which pure strategy to commit to could not result in a
utility greater than $2$ for player $1$ in the game above.
 
Because the game tree has infinitely many nodes, it cannot be explicitly
written down on paper or---perhaps more importantly---in computer memory.
An algorithm for computing the optimal mixed strategy to commit to must
operate on a different representation of the game---most naturally, the
original normal form from which the game was obtained.  Of course, an
alternative is to discretize the space of mixed strategies, choosing only a
finite subset of them to stand in as ``representatives'' in the hope of
getting a reasonable approximation.  This, however, gives up on exactly
representing the game, and moreover is not even a computationally efficient
way of solving the game, as we will discuss in more detail later.  A
closely related issue is that this infinitely-sized extensive-form
representation does little to facilitate seeing the underlying structure of
the game.\footnote{\cite{Schelling60:Strategy} similarly suggests that, by
  the time we have incorporated aspects such as commitment moves into a
  standard game-theoretic representation of the game at hand, we have
  abstracted away these issues and are at some level not really studying
  them anymore.}  From seeing it (or a finite approximation of it), the
viewer may not even realize that player $1$'s actions in the game
correspond to the set of all mixed strategies of an original normal-form
game.\footnote{It is easy to be misled by Figure~\ref{fi:mixed_extensive}
  into thinking that it does make this fairly obvious, due to the natural
  ordering of the mixed strategies from left to right.  However, this is an
  artifact of the fact that there are only two pure strategies for player 1
  in the original game.  If there were three pure strategies, it would not
  be possible to order the mixed strategies so naturally from left to
  right.  We could in principle visualize the resulting tree in three
  dimensions instead of two.  For more pure strategies, this of course
  becomes problematic.  More importantly, such visualizations are
  technically {\em not} part of the extensive form.  The extensive form
  only specifies a {\em set} of actions for each node, and no ordering,
  distance function, or topology on them.  Such are only added when we draw
  the tree on a piece of paper (or in three dimensions, etc.).}

Still, one may argue that, while it may be true that the extensive form
obscures some of the structure of the game, this is not sufficient reason
to study Stackelberg mixed strategies separately (i.e., as directly
providing a solution for a game represented in normal form).  After all, it
is often the case that, when we consider a solution concept in the special
context of some specific family of games, additional structure appears that
was not there in the general case.  However, in what follows, we will see
that there are other reasons to study Stackelberg mixed strategies
separately.

\section{Von Neumann's heritage: Zero-sum games}
\label{se:neumann}

If there is one class of games that game theory can be said to truly {\em
  solve} (other than games solvable by iterated dominance), it is that of
two-player zero-sum games.  This is due to von Neumann's famous {\em
  minimax theorem}~\citep{Neumann28:Zur}.  In such games, there are two
players with pure strategy sets $S_1$ and $S_2$, respectively, and for all
$s_1 \in S_1, s_2 \in S_2$, we have
$u_1(s_1,s_2)+u_2(s_1,s_2)=0$.\footnote{{\em Constant-sum} games, in which
  $u_1(s_1,s_2)+u_2(s_1,s_2)=c$ for some constant $c$, are effectively
  equivalent.}  Consider the scenario where player $1$ is extremely
conservative and assumes that, no matter which mixed strategy she chooses,
player $2$ will manage to choose the strategy that is worst for player $1$.
Under this pessimistic assumption, player $1$ can still guarantee herself
$$\max_{\sigma_1 \in \Sigma_1} \min_{s_2 \in S_2} u_1(\sigma_1,s_2)$$ where
$\Sigma_1=\Delta(S_1)$ is the set of player $1$'s mixed strategies.
Strategies $\sigma_1$ that achieve this maximum are known as {\em maximin}
strategies.  If player $2$ were to make a similar pessimistic assumption,
he could guarantee himself
$$\max_{\sigma_2 \in \Sigma_2} \min_{s_1 \in S_1} u_2(s_1,\sigma_2)$$
Because the game is zero-sum, instead of trying to maximize his own
utility, player $2$ could equivalently try to minimize player $1$'s
utility.  Then, a pessimistic player $2$ could guarantee that player $1$
gets no more than
$$\min_{\sigma_2 \in \Sigma_2} \max_{s_1 \in S_1} u_1(s_1,\sigma_2)$$
Strategies $\sigma_2$ that achieve this minimum are known as {\em minimax}
strategies.
Indeed, note that $$\min_{\sigma_2 \in \Sigma_2} \max_{s_1 \in S_1}
u_1(s_1,\sigma_2) = - \max_{\sigma_2 \in \Sigma_2} \min_{s_1 \in S_1}
u_2(s_1,\sigma_2)$$ but this is not yet the minimax theorem.  Rather, the
minimax theorem states that $$\max_{\sigma_1 \in \Sigma_1} \min_{s_2 \in
  S_2} u_1(\sigma_1,s_2) = \min_{\sigma_2 \in \Sigma_2} \max_{s_1 \in S_1}
u_1(s_1,\sigma_2)$$  This quantity is known as the {\em value} of the game.
If (for example) the game is played repeatedly by sophisticated players, it
seems very reasonable to expect that this is the average value that
player $1$ will obtain from a round of play over time.  If she were getting
less, she should just switch to a strategy that guarantees at least the
value.  If she were getting more, then player $2$ should switch to a
strategy that guarantees that player $1$ gets at most the value.

From the minimax theorem, it is straighforward to deduce that a strategy
profile is a Nash equilibrium of a two-player zero-sum game if and only if
player $1$ plays a maximin strategy and player $2$ plays a minimax
strategy.  Hence, the concept of Nash equilibrium provides a generalization
of these strategies to general-sum games.  On the other hand, it is even
easier to see that the Stackelberg mixed strategies for player $1$ coincide
with her maximin strategies in a two-player zero-sum game; the definition
of a Stackelberg mixed strategy is a straightforward generalization of that
of a maximin strategy in such games.  Hence, Stackelberg mixed strategies
and Nash equilibrium strategies coincide in two-player zero-sum games.
This should not be surprising, because any solution concept that does not
coincide with (or refine) maximin/minimax strategies in two-player zero-sum
games would seem suspect given the minimax theorem.  Nevertheless,
Stackelberg mixed strategies and Nash equilibrium strategies generalize to
general-sum games in different ways, and arguments can be given both ways
as to which is more natural.  But viewing Stackelberg mixed strategies
(only) as the solution to an extensive-form game obscures this and would
appear to leave Nash equilibrium (or related equilibrium concepts) as the
only generalization.

We will return to properties that are obscured by not studying Stackelberg
mixed strategies directly on the normal form in Section~\ref{se:obscure}.
First, however, we will consider computational aspects.

\section{The computational angle}

Historically, the development of our understanding of the minimax theorem
was tied up with the development of {\em linear programming}.  A linear
program describes an optimization problem over multiple variables, with
multiple linear inequality constraints on these variables as well as an
objective to be minimized or maximized.  Figure~\ref{fi:minimax_LP} shows
how the problem of finding a maximin strategy can be formulated as a linear
program (as is well known).
\begin{figure}
\begin{center}
{\normalsize
\fbox{
  \parbox{4.53in}{
    maximize $v_1$\\
    subject to\\
    $(\forall s_2 \in S_2) \ v_1 - \sum_{s_1 \in S_1} u_1(s_1,s_2) p_{s_1} \leq 0$\\
    $\sum_{s_1 \in S_1} p_{s_1} = 1$\\
    $(\forall s_1 \in S_1) \ p_{s_1} \geq 0$ }}}
\end{center}
\caption{Linear program formulation for computing a maximin strategy for
  player $1$. $p_{s_1}$ is a variable indicating the probability placed on
  player $1$'s pure strategy $s_1$. The first constraint requires that
  $v_1$ be at most the utility that player $1$ gets when player $2$
  best-responds, and the goal is to maximize this minimum utility for
  player $1$.}
\label{fi:minimax_LP}
\end{figure}
\cite{Dantzig51:Proof} showed that, from a computational viewpoint, the two
problems are equivalent.\footnote{In fact, he pointed out that there was a
  case in which his reduction from linear programs to zero-sum games does
  not work; this gap was later filled by~\cite{Adler13:Equivalence}.}

Besides providing a mathematically elegant way to model many optimization
problems, linear programs are useful for determining the {\em computational
  complexity} of problems.  An example of a computational problem is that
of finding maximin strategies of two-player zero-sum games, represented in
normal form (and any specific two-player zero-sum game would be an {\em
  instance} of this problem).  Computer scientists design {\em algorithms}
for solving such problems.  Such an algorithm is generally required to
provide the correct output for {\em any} input---e.g., any two-player
zero-sum game.  With some training, designing correct algorithms is usually
not that hard; however, for many problems, designing {\em fast} algorithms
is challenging.  One may wonder why we should really care whether
algorithms are fast.  So what if my computer needs to work a little harder?
I can wait a few seconds if needed.  The flaw in this reasoning is that for
many problems, the runtime of the obvious algorithms scales {\em
  exponentially} in the size of the input, so that as we increase the size
of the problem instances, rather quickly we find instances that would take
the algorithm more than the lifetime of the universe to solve, even on the
fastest computer available.  In contrast, other algorithms have the
property that their runtime scales only as a {\em polynomial} function in
the size of the input.  Problems for which such algorithms exist are
generally considered {\em tractable}, and the algorithm is said to be {\em
  efficient}.  Note that the same problem may have two correct algorithms,
one of which scales exponentially and one of which (perhaps one that
requires more design effort) scales polynomially; in this case, still, the
problem is considered tractable.  (It is always possible to find a slow
algorithm for a problem; the question is whether fast ones exist.)
 
It is known that linear programs can be solved in polynomial
time~\citep{Khachiyan79}.  That means that any problem that can be
rephrased as (or, technically, {\em reduced to}) a linear program can also
be solved in polynomial time. (Note that this does require that the linear
program itself can be obtained in polynomial time, and {\em a fortiori}
that the linear program has polynomial size---an exponentially sized linear
program could not be written down in polynomial time.)  In particular, this
implies that the problem of computing a maximin strategy of a two-player
zero-sum game can be solved in polynomial time.

Now, what about the more general problem of computing a Nash equilibrium of
a two-player (general-sum) game represented in normal form?  This one turns
out to be significantly trickier.  There is no known linear program
formulation for this problem, and more generally, no polynomial-time
algorithms are known.  Perhaps the best-known algorithm---the {\em
  Lemke-Howson} algorithm~\citep{Lemke64:Equilibrium}---is known to require
exponential time on some families of games~\citep{Savani04:Exponentially}.
(Other algorithms more obviously require exponential time in some
cases~\citep{Dickhaut91:Program,Porter04:Simple,Sandholm05:Mixed}.)  Can we
prove it is in fact impossible to design a polynomial-time algorithm for
this problem?  As is the case for many other computational problems, we do
not currently have the techniques to unconditionally prove this.  What
computer scientists often can do in these cases is to prove the following
type of result: ``If this problem can be solved in polynomial time, then so
can any problem in the class $C$ of problems.''  In this case, the original
problem is said to be $C$-hard (and, if the problem additionally is itself
a member of $C$, it is said to be $C$-complete).  The most famous such
class is NP. Indeed, problems such as the following turn out to be
NP-complete: ``Given a two-player game in normal form, determine whether it
has a Nash equilibrium in which pure strategy $s_1$ receives positive
probability,'' or ``Given a two-player game in normal form, determine
whether it has a Nash equilibrium in which the sum of the players' expected
utilities exceeds a threshold
$\epsilon$''~\citep{Gilboa89:Nash,Conitzer03:Nash}.  For the problem of
computing just one Nash equilibrium of a two-player game in normal
form---i.e., any one Nash equilibrium will do---the problem is known to be
PPAD-complete~\citep{Daskalakis09:Complexity,Chen09:Settling}.\footnote{\cite{Papadimitriou94:On}
  introduced the class PPAD.  \cite{Daskalakis05:Three} showed that the
  problem is PPAD-hard for three players; \cite{Chen05:Settling} then
  obtained the stronger result that it is PPAD-hard even for two players.
  \cite{Etessami10:Complexity} proved that with three or more players, the
  problem of computing an exact Nash equilibrium, rather than an
  $\epsilon$-equilibrium, is FIXP-complete.}  The precise definition of
these classes is not of importance here; suffice it to say that computer
scientists generally give up on designing an efficient algorithm for the
problem when such a complexity result is found for it.

Then, what about computing a Stackelberg mixed strategy for a two-player
game represented in normal form?  One approach---arguably the most natural
one when we do not study Stackelberg mixed strategies separately---would be
to convert the game to the extensive-form representation of the leadership
model, and solve the resulting game for a subgame perfect Nash equilibrium.
As discussed before, one problem with this approach is that the extensive
form of such a game in fact has infinite size, and can therefore not be
(directly) represented in computer memory.  A natural (though only
approximate) approach is to discretize the space of distributions to which
player $1$ can commit.  For any $N$, we can restrict our attention to the
finitely many distributions that only use probabilities that are multiples
of $1/N$.  However, this still results in $N + |S_1| - 1 \choose |S_1|-1$
different distributions.  (This is equal to the number of ways in which $N$
indistinguishable balls---corresponding to the $N$ atomic units of $1/N$
probability mass---can be placed in $S_1$ distinguishable
bins---corresponding to the different pure strategies for player $1$.)
This number is exponential in the number of pure strategies for player $1$,
so this approach cannot lead us to a polynomial-time algorithm (even
ignoring the fact that it in general will not provide an exact solution).

As it turns out, though, it is in fact possible to solve this problem in
polynomial time, if we avoid converting the game into extensive form first.
Because computing a Stackelberg mixed strategy is a generalization of
computing a maximin strategy in a two-player zero-sum game, it should not
come as a surprise that this algorithm relies on linear programming.  The
algorithm uses a divide-and-conquer approach, as follows.  For every pure
strategy $s_2^* \in S_2$ for player $2$, we ask the following question: (Q)
what is the highest utility that player $1$ can obtain, {\em under the
  condition that player $1$ plays a mixed strategy $\sigma_1$ to which
  $s_2^*$ is a best response} (and assuming that player $2$ in fact
responds with $s_2^*$)?  For some strategies $s_2^*$, it may be the case
that there is {\em no} $\sigma_1$ to which $s_2^*$ is the best response,
and this will correspond to the linear program having no feasible
solutions---but this obviously cannot be the case for {\em all} of player
$2$'s strategies.  Among the ones that do have feasible solutions, we
choose one that gives the highest objective value, and the corresponding
mixed strategy $\sigma_1$ is an (optimal) Stackelberg mixed strategy for
player $1$.  It remains to be shown how to formulate (Q) as a linear
program.  This is shown in Figure~\ref{fi:standard_LP}. (Later on, I will
discuss another formulation for the problem, as a single linear program
(Figure~\ref{fi:correlated_commit}).)  The main point to take away is that
the extensive-form view of Stackelberg mixed strategies does little to lead
us to an efficient algorithm for computing them, whereas studying these
strategies separately, as providing a solution for games represented in
normal form, suggests that a linear programming approach may succeed, which
in fact it does.
\begin{figure}
\begin{center}
{\normalsize \fbox{
    \parbox{4.53in}{
      maximize $\sum_{s_1 \in S_1} u_1(s_1,s_2^*) p_{s_1}$\\
      subject to\\
      $(\forall s_2 \in S_2) \ \sum_{s_1 \in S_1} (u_2(s_1,s_2^*) -
      u_2(s_1,s_2))
      p_{s_1} \geq 0$\\
      $\sum_{s_1 \in S_1} p_{s_1} = 1$\\
      $(\forall s_1 \in S_1) \ p_{s_1} \geq 0$ }}}
\end{center}
\caption{Linear program formulation for computing a Stackelberg mixed
  strategy (more precisely, an optimal strategy for player $1$ that induces
  $s_2^*$ as a best
  response)~\citep{Conitzer06:Computing,Stengel10:Leadership}.  $p_{s_1}$
  is a variable indicating the probability placed on player $1$'s pure
  strategy $s_1$.  The objective gives player 1's expected utility given
  that player $2$ responds with $s_2^*$, and the first constraint ensures
  that $s_2^*$ is in fact a best response for player $2$.}
\label{fi:standard_LP}
\end{figure}

\section{Other properties that are easier to interpret when studying
  Stackelberg mixed strategies separately}
\label{se:obscure}

As discussed in Section~\ref{se:neumann}, if we do not separately study how
Stackelberg mixed strategies provide solutions to 2-player normal-form
games, this obscures that they are a generalization of maximin strategies.
In this section, I discuss some other properties of Stackelberg mixed
strategies that involve comparisons to Nash equilibria of the
simultaneous-move game.  I argue that it is easier to get insight into
these properties if we do study Stackelberg mixed strategies separately.

One may wonder about the following: is commitment always advantageous,
relative to, say, playing a Nash equilibrium of the simultaneous-move game?
It is clear that committing to a {\em pure} strategy is not always a good
idea.  For example, when playing Rock-Paper-Scissors, presumably it is not
a good idea to commit to playing Rock and make this clear to your opponent.
On the other hand, committing to the mixed strategy $(1/3,1/3,1/3)$ does
not hurt one bit.\footnote{An exception is, of course, if we play against
  an exploitable non-game-theoretic player, such as one who always plays
  Scissors.}  More generally, in any two-player zero-sum game, (optimally)
committing to a mixed strategy beforehand does not hurt (or help) one bit:
this is exactly what the minimax theorem tells us.  But what about in
general-sum games?  We have already seen that it can (strictly) help
there,\footnote{For a study of {\em how much} it can help,
  see~\cite{Letchford10:Value}.}  but does it ever hurt?  It turns out that
it does not, and it is not hard to get some intuition why.  Consider any
Nash equilibrium $(\sigma_1,\sigma_2)$ of the simultaneous-move game---for
example, one that is optimal for player $1$.  Then, if player $1$ commits
to playing $\sigma_1$, then any one of the pure strategies in $\sigma_2$'s
support is a best response.  If we assume that player $2$ breaks ties in
player $1$'s favor---or if the game is such that player $1$ can, by
adjusting her strategy slightly, make the most favorable strategy in
$\sigma_2$'s support the unique best response for player $2$---then player
$1$ must be at least as well off as in the case where player $2$ responds
with $\sigma_2$, which is the Nash equilibrium case.  Without these
assumptions, things become significantly hairier, because depending on how
player $2$ breaks ties, player $1$ may end up with any utility in an
interval---but it can be shown that this interval is still more favorable
than the corresponding interval for Nash
equilibrium~\citep{Stengel10:Leadership}.  At least in my view, comparisons
such as these are more natural when we study Stackelberg mixed strategies
separately, so that we are comparing player $1$'s utility in two different
solutions of the same game, rather than comparing player $1$'s utility
across two different games.

As another example,~\cite{Kiekintveld09:Computing} introduce a class of
games called {\em security games}, which involve a defender and an
attacker.  In these games, the defender chooses how to allocate its
resources to (subsets of) the targets, and the attacker chooses a target to
attack.  Both players' utilities are a function of (1) which target is
attacked and (2) whether that target is defended by some resource(s).
Holding the attacked target fixed, the defender prefers for it to be
defended, and the attacker prefers for it not to be defended.
\cite{Korzhyk11:Stackelberg} show that, under a minor assumption---namely,
that if a resource can (simultaneously) defend a given set of targets, then
it can also defend any subset of that set---every Stackelberg mixed
strategy for the defender is also a Nash equilibrium strategy for the
defender (in the simultaneous-move version of the game).  (Moreover, it is
shown that the Nash equilibria of the simultaneous-move game satisfy {\em
  interchangeability}: if $(\sigma_1,\sigma_2)$ and $(\sigma_1',\sigma_2')$
are equilibria, then necessarily so are $(\sigma_1,\sigma_2')$ and
$(\sigma_1',\sigma_2)$.)  Hence, in a sense, for these games, Stackelberg
mixed strategies are a refinement of Nash equilibrium strategies (for the
defender).  Now, the point here is not to place undue emphasis on security
games.  Rather, the point is, again, that this type of refinement property
is very cumbersome to state if we strictly hold to the view that
Stackelberg mixed strategies are just subgame perfect Nash equilibria of a
different game.  We would have to make a statement about how the solutions
to two different games relate to each other, rather than just being able to
state that one concept is a refinement of the other.

\section{The analogous (and related) case of correlated equilibrium}

Unlike Stackelberg mixed strategies, {\em correlated
  equilibrium}~\citep{Aumann74:Subjectivity} is commonly considered a
solution concept in its own right.  Roger Myerson has been quoted as saying
that: ``If there is intelligent life on other planets, in a majority of
them, they would have discovered correlated equilibrium before Nash
equilibrium.''  In fact, I personally believe that most of them would have
discovered Nash equilibrium before correlated equilibrium, like we did, but
this will be a difficult one to settle.  The point, anyway, is well taken:
correlated equilibrium is a natural solution concept that is technically
more elegant than Nash equilibrium in a number of ways.  Having thus built
up the suspense, let us now define the correlated equilibrium concept.

In a correlated equilibrium of a $2$-player\footnote{All of this is easily
  generalized to $n$ players, but for simplicity I will stick to two
  players here.}  game, an ordered pair of signals
$(\theta_1,\theta_2) \in \Theta_1 \times \Theta_2 = \Theta$ is drawn
according to some distribution $p: \Theta \rightarrow [0,1]$.  (Note that
the $\theta_i$ need not be independent or identically distributed.)  Each
player $i$ receives her signal $\theta_i$, and based on this takes an
action in the game.  That is, player $i$ has a strategy
$\tau_i: \Theta_i \rightarrow \Sigma_i$, where $S_i$ is the set of actions
for player $i$ in the game and $\Sigma_i = \Delta(S_i)$ is the set of
probability distributions over these actions.\footnote{The notation here is
  a bit nonstandard: in isolation, it would be more natural to use $A_i$ to
  denote the set of actions and $S_i$ to denote the set of pure strategies,
  i.e., mappings from signals to actions.  However, in order to make the
  comparison to other concepts easier, it will help to stick to using $s_i$
  for the rows and columns of the game.}  All of this collectively
constitutes a correlated equilibrium if it is optimal for each player to
follow her strategy assuming that the other does so as well.  That is, for
every player $i$ and every signal $\theta_i$ that $i$ receives with
positive probability
($p(\theta_i) = \sum_{\theta_{-i}} p(\theta_i,\theta_{-i}) >0$), and for
every action $s_i$ that player $i$ might take, we have
$$\sum_{\theta_{-i}} p(\theta_{-i} | \theta_i)
(u_i(\tau_i(\theta_i),\tau_{-i}(\theta_{-i})) -
u_i(s_i,\tau_{-i}(\theta_{-i}))) \geq 0$$
That is, the strategies $\tau_i$ are an equilibrium of the Bayesian game
defined by the distribution over the signals. (Note, however, that this
distribution is considered part of the solution.)

It is well known and straightforward to show that, if all we care about is
the resulting distribution over outcomes $S$---where $S = S_1 \times S_2$
and the probability of an outcome $s = (s_1,s_2)$ is
$$P(s_1,s_2) = \sum_{(\theta_1,\theta_2) \in \Theta} p(\theta_1,\theta_2)
\tau_1(\theta_1)(s_1) \tau_2(\theta_2)(s_2)$$
where $\tau_i(\theta_i)(s_i)$ is the probability that the distribution
$\tau_i(\theta_i)$ places on $s_i$---then it is without loss of generality
to
\begin{itemize}
\item let each player's signal space coincide with that player's action
  space, i.e., $\Theta_i = S_i$,
\item consider the strategies where players simply follow their signals,
  i.e., if $\theta_i=s_i$, then $\tau_i(\theta_i)$ is the distribution that
  places probability $1$ on $s_i$, and
\item (consequently) for $(\theta_1,\theta_2)=(s_1,s_2)$, we have
  $P(s_1,s_2)=p(\theta_1,\theta_2)$.
\end{itemize}
That is, if a correlated equilibrium resulting in probability distribution
$P$ over outcomes $S$ exists, then it is also a correlated equilibrium to
draw the outcome directly according to that distribution and signal to each
player the action that she is supposed to play in this outcome (but nothing
more).  Hence, we may dispense with the $\theta_i$ notation.  It also
allows us to describe the set of correlated equilibria with the set of
inequalities in Figure~\ref{fi:correlated}.

\begin{figure}
\begin{center}
{\normalsize
\fbox{
  \parbox{4.53in}{
    $(\forall s_1, s_1' \in S_1) \ \sum_{s_2 \in S_2} (u_1(s_1,s_2) -
    u_1(s_1',s_2)) p_{s_1,s_2} \geq 0$\\
    $(\forall s_2, s_2' \in S_2) \ \sum_{s_1 \in S_1} (u_2(s_1,s_2) -
    u_2(s_1,s_2')) p_{s_1,s_2} \geq 0$\\
    $\sum_{s_1 \in S_1} \sum_{s_2 \in S_2} p_{s_1,s_2} = 1$\\
    $(\forall s_1 \in S_1, s_2 \in S_2) \ p_{s_1,s_2} \geq 0$ }}}
\end{center}
\caption{Linear inequalities specifying the set of correlated equilibria of
  a 2-player game (this can easily be generalized to $n$-player games).
  $p_{s_1,s_2}$ is a variable representing the probability of the profile
  $(s_1,s_2)$ being played.  The first constraint says that player $1$,
  upon receiving a signal to play $s_1$, should not be better off playing
  another $s_1'$ instead.  The second constraint, for player $2$, is
  similar.}
\label{fi:correlated}
\end{figure}

Now let us consider the question of whether correlated equilibrium
``deserves'' to be considered a solution concept in its own right.  One
might argue that it does not, in a way that is analogous to the argument
against studying Stackelberg mixed strategies separately, as follows.  We
can define the set of correlated equilibria of a game $G$ simply as the set
of all Nash equilibria\footnote{Nash equilibria of a game with private
  information are often referred to as {\em Bayes-Nash} equilibria.}  of
all games that result from extending $G$ with signals to the players, as
described at the beginning of this section.  Hence, the concept can be seen
as derivative rather than primitive.  Instead of thinking about it as a
separate solution concept, we can simply think of it as the application of
the Nash equilibrium concept to a modified game (the game extended with
signals).\footnote{It could be argued that the analogy is imperfect because
  in the Stackelberg version of the argument, the game is modified to a
  {\em single} (two-stage) game, whereas in the correlated equilibrium
  version of the argument, two different correlated equilibria potentially
  require different ways of modifying the game, extending them with
  different signaling schemes.  It is not entirely clear to me how
  significant this distinction is.  In any case, if two correlated
  equilibria require different signaling schemes, then consider a new,
  joint signaling scheme where each player receives the signals from {\em
    both} signaling schemes, with the signals drawn independently across
  the two schemes.  Then, both correlated equilibria are (Bayes-)Nash
  equilibria of the game with the joint signaling scheme (with the players
  simply ignoring the part of the signal that corresponds to the other
  equilibrium).  Taking this to the limit, we may imagine a single,
  universal signaling scheme such that all correlated equilibria of
  interest are Nash equilibria of the game with this universal signaling
  scheme.}

Of course, my aim here is not to {\em actually} argue that correlated
equilibrium should not be considered a separate solution concept.
Correlated equilibria have many elegant and useful properties that would be
obscured by thinking of them merely as the application of the Nash
equilibrium concept to an enriched game.  The fact that correlated
equilibria can be computed in polynomial time using the linear feasibility
formulation in Figure~\ref{fi:correlated} is one example of this: an
explicit Bayesian game formulation (with signals) would presumably not be
helpful for gaining insight into this polynomial-time computability, as
such a formulation would involve exponentially many strategies.  Rather,
the point is that the case for studying correlated equilibrium separately
is, in my view, quite similar to the case for studying Stackelberg mixed
strategies separately.

Indeed, returning to the line of reasoning from Section~\ref{se:obscure},
when {\em both} correlated equilibria and Stackelberg mixed strategies are
studied in their own right---as applying directly to the normal form of the
game, rather than being the solutions to two different games---it becomes
apparent that they are in fact closely related.  Consider again the linear
program in Figure~\ref{fi:standard_LP}, which is used to compute an optimal
strategy for the leader under the constraint that a particular pure
strategy for the follower must be optimal, so that solving this linear
program for every pure follower strategy gives an optimal solution.
\cite{Conitzer11:Commitment} observe that we can combine all these linear
programs (one for each pure follower strategy) into a larger single linear
program. The resulting linear program is given in
Figure~\ref{fi:correlated_commit}.
\begin{figure}
\begin{center}
{\normalsize
\fbox{
  \parbox{4.53in}{
    maximize $\sum_{s_1 \in S_1} \sum_{s_2 \in S_2} u_1(s_1,s_2) p_{s_1,s_2}$\\
    subject to\\
    $(\forall s_2, s_2' \in S_2) \ \sum_{s_1 \in S_1} (u_2(s_1,s_2) -
    u_2(s_1,s_2')) p_{s_1,s_2} \geq 0$\\
    $\sum_{s_1 \in S_1} \sum_{s_2 \in S_2} p_{s_1,s_2} = 1$\\
    $(\forall s_1 \in S_1, s_2 \in S_2) \ p_{s_1,s_2} \geq 0$ }}}
\end{center}
\caption{A single linear program for computing a Stackelberg mixed
  strategy~\citep{Conitzer11:Commitment}.  This linear program can be
  obtained by combining the linear programs (one for each $s_2$) from
  Figure~\ref{fi:standard_LP} and renaming the variable $p_{s_1}$ from the
  linear program corresponding to $s_2$ to $p_{s_1,s_2}$.  An optimal
  solution for which there exists some $s_2^*$ such that $p_{s_1,s_2} = 0$
  whenever $s_2 \neq s_2^*$ is guaranteed to exist.}
\label{fi:correlated_commit}
\end{figure}
The constraints of this linear program are exactly the set of linear
inequalities above for correlated equilibrium (Figure~\ref{fi:correlated}),
except that only the constraints for player $2$ appear.  Moreover, the
objective is to maximize player $1$'s utility.  An immediate corollary of
this is a result (which was earlier proved directly
by~\cite{Stengel10:Leadership}) that a Stackelberg mixed strategy is at
least as good for the leader as any correlated equilibrium, because if we
add the constraints for player $1$ we get a linear program for finding the
best correlated equilibrium for player $1$---and adding constraints can
never improve the optimal value of a linear program.  One way to interpret
the linear program in Figure~\ref{fi:correlated_commit} is as follows:
player $1$ now gets to commit to a {\em correlated} strategy, where she
chooses a profile $(s_1,s_2)$ according to some distribution, signals to
player $2$ which action $s_2$ he should play (where there is a constraint
on the distribution such that player $2$ is in fact best off listening to
this advice), and plays $s_1$ herself.  \cite{Conitzer11:Commitment} prove
that there always exists an optimal solution where player $1$ always sends
the same signal $s_2^*$ to player $2$, so that effectively player $1$ is
just committing to a mixed strategy.  (When there are $3$ or more players,
then the optimal solution may require true correlation.)  Again, the main
point is that this close relationship between Stackelberg mixed strategies
and correlated equilibrium is obscured if we think of Stackelberg mixed
strategies in terms of extensive-form games (or, for that matter, if we
think of correlated equilibrium in terms of Bayesian games).

\section{Conclusion}

In game theory, sometimes one solution concept is equivalent to the
application of another solution concept to a modified representation of the
game.  In such cases, is it worthwhile to study the former in its own
right, as it applies to the original representation?  It appears difficult
to answer this question in general, without knowing either what the
solution concept is or what the context is in which we are attempting to
answer the question.  In this article, I have investigated this question
for the specific concept of Stackelberg mixed strategies.  Often, game
theorists think of Stackelberg models as just that---a different model of
how the game is to be played, rather than a different way of solving the
game.  There are certainly good reasons for this view.  However, my overall
conclusion is that, in the context of Stackelberg {\em mixed} strategies,
limiting oneself to this view comes at too great a cost.  Studying them in
their own right, as providing solutions of normal-form games, often
facilitates mathematical analysis---making connections to other concepts
such as minimax strategies, Nash equilibrium, and correlated equilibrium
more apparent---as well as computational analysis, allowing one to find
efficient direct algorithms rather than attempting to work with
discretizations of infinitely sized objects.

I should emphasize, however, that the possibility of viewing these
strategies as solutions of an extensive-form game surely remains valuable
too.  For example, from the perspective of epistemic game theory,
Stackelberg mixed strategies may be easiest to justify via this
interpretation.  Similarly, I would argue that both views are valuable for
correlated equilibrium, which I have argued is an analogous case: while it
is extremely useful for mathematical and computational purposes to study
correlated equilibrium as a solution concept for normal-form games in its
own right, as indeed it usually is viewed, seeing it as a (Nash)
equilibrium of an enriched game has its own benefits---not the least of
which is that this is a common and natural way to introduce the concept.
Hence, I believe that the choice between the two views is much like the
choice between seeing the young woman and the old woman in the famous
ambiguous image.  While we generally cannot hold both views simultaneously,
if we do not allow our minds to switch from one view to the other, we miss
out on much of what is there.

\noindent {\bf Acknowledgments.}
I thank the LOFT and Synthese reviewers for helpful feedback, and ARO and
NSF for support under grants W911NF-12-1-0550, W911NF-11-1-0332,
IIS-0953756, CCF-1101659, CCF-1337215, and IIS-1527434.

\bibliographystyle{plainnat}
\bibliography{shouldstackbeseparatesolutionconcept_synthese.bib}

\end{document}